\documentclass[aps,twocolumn,showpacs,superscriptaddress,amsmath,amssymb,amsfonts,floatfix,longbibliography]{revtex4-1}

\usepackage[T1]{fontenc} 
\usepackage{graphicx}
\usepackage{float}
\usepackage{todonotes,soul}
\usepackage{dcolumn}
\usepackage{bm}
\usepackage{amssymb}
\usepackage{microtype}
\usepackage{xfrac}
\usepackage{gensymb}
\usepackage[most]{tcolorbox}
\usepackage{xcolor}
\usepackage{multirow}
\usepackage{enumitem}
\usepackage{natbib,hyperref}
\usepackage{graphicx}
\usepackage{booktabs}
\usepackage{chemformula}
\usepackage{array}

\hypersetup{
	colorlinks=true, 
	citecolor=blue,  
}

\begin{document}

\title{Geometric Approach to the High-Throughput Identification of Honeycomb Materials}

\author{Lex M. Rouquette}
    \affiliation{Stewart Blusson Quantum Matter Institute, University of British Columbia, Vancouver, BC V6T 1Z4, Canada}
    \affiliation{Department of Physics \& Astronomy, University of British Columbia, Vancouver, BC V6T 1Z1, Canada}

\author{Alannah M. Hallas}
\email[Email: ]{alannah.hallas@ubc.ca}
    \affiliation{Stewart Blusson Quantum Matter Institute, University of British Columbia, Vancouver, BC V6T 1Z4, Canada}
    \affiliation{Department of Physics \& Astronomy, University of British Columbia, Vancouver, BC V6T 1Z1, Canada}
    \affiliation{Canadian Institute for Advanced Research (CIFAR), Toronto, ON, M5G 1M1, Canada}

\date{\today}

\begin{abstract}
The honeycomb lattice is an iconic structural motif in condensed matter physics. However, emerging theoretical models require the identification of new honeycomb lattice materials with strict specification on their structural, magnetic, and electronic characteristics. In this work, we introduce a filtering algorithm using simple geometric arguments to detect materials with honeycomb structural motifs. We classify four unique categories of lattices: conventional honeycombs, hyperhoneycombs, staircase or zigzag honeycombs, and fully 3D honeycombs. We also identify several distortion motifs in the conventional honeycomb set: elongated, sheared, and buckled honeycombs. We describe several screening criteria for further analysis of honeycomb systems, specifically the quantification of exfoliability in 2D materials, the identification of magnetic honeycombs, and a density-of-states peak quality metric for identification of electronic instabilities. The database containing the results of the filters and subsequent analysis is a useful tool for identifying novel materials for targeted physical phenomena for experimental characterization.

\end{abstract}

\maketitle

\section{\label{sec:Introduction}Introduction}

Honeycomb motifs, which derive their name from the wax cells formed by bees to store honey, are one of the most recognizable patterns in nature. The prevalence of honeycomb patterns can, in part, be understood as a direct consequence of the ``honeycomb conjecture'', which states that of all possible equal area tilings of a two-dimensional plane, honeycombs minimize the total perimeter, a result that was mathematically proven in 1999 by Thomas C. Hales~\cite{hales2001honeycomb}. As a consequence of this efficiency, honeycombs achieve excellent strength-to-weight ratios and are commonly incorporated in human-engineered materials, particularly those with aerospace applications. At the nanoscale, honeycomb motifs are also commonly found in the atomic arrangements of crystalline materials, where they are often referred to as the honeycomb lattice. This terminology is in fact a misnomer, as the honeycomb is not itself a lattice in the technical sense but is instead formed by tiling two symmetrically inequivalent atomic positions onto a hexagonal lattice.

\begin{figure*}
    
    \centering
    \includegraphics[width=0.9\linewidth]{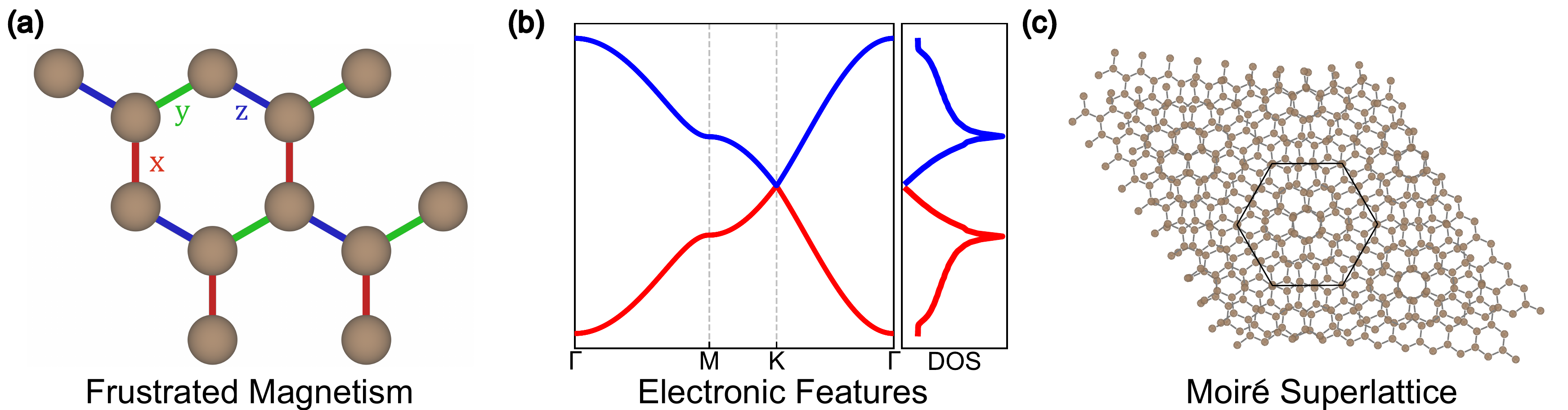}
    \caption{(a) The Kitaev model of magnetism identifies three different interactions between honeycomb sites, denoted by $x,y$, and $z$. It is an analytically solvable model with a quantum spin liquid as its ground state. (b) Honeycomb lattices have two electronic bands in the tight-binding model, producing van Hove singularities around the M point, and a Dirac cone around K. These bands split with distortions or other effects that alter the hopping between the honeycomb sites. (c) Placing honeycomb monolayers on top of each other with a slight rotation produces a Moir\'e pattern. Some of these twisted bilayer materials have been shown to exhibit superconductivity \cite{bilayergraphene}.}
    \label{fig:intro_fig}
\end{figure*}

Honeycomb materials have risen to a special prominence in condensed matter physics following the 2004 isolation of graphene \cite{graphene_original}, a material composed of a single atomic layer of carbon arranged in a honeycomb network. Among its many remarkable properties, graphene hosts electronic excitations (Fig. \ref{fig:intro_fig}b) that behave as massless Dirac fermions \cite{dirac_fermions}, arising from symmetry-protected band crossings in its electronic structure, which give rise to exceptionally high carrier mobilities \cite{carrier_mobility} and unconventional quantum Hall phenomena \cite{quantum_hall_effect}. In addition to these electronic properties, graphene exhibits notable mechanical characteristics, combining exceptional tensile strength and stiffness with the flexibility expected of an atomically thin material \cite{graphene_mech}. 
More recently, advances in the assembly of van der Waals heterostructures have revealed that the relative stacking and twisting of honeycomb layers (Fig \ref{fig:intro_fig}c) can dramatically modify their electronic properties, leading to flat electronic bands \cite{graphene_flat}, strongly correlated insulating states \cite{graphene_correlated}, superconductivity \cite{bilayergraphene}, and topological phases \cite{graphene_topology}. 

Leaving the purely two-dimensional limit, there has also been significant interest in honeycomb lattices occupied by magnetic species and embedded in structurally three-dimensional materials. In these cases, provided that the atomic spacing within the honeycomb layer is smaller than the out of plane spacings, the magnetic interactions can have a highly two-dimensional character. It is worth emphasizing that honeycombs are not intrinsically geometrically frustrated: their bipartite lattices can support simple ``up-down-up-down'' antiferromagnetic motifs. However, with the addition of further neighbor exchange and anisotropic couplings, the honeycombs can quickly depart from simple magnetic orders and enter regimes of strongly competing interactions \cite{honeycomb_stronginteraction}. The effects of this frustration are further enhanced by the low connectivity of the honeycomb lattice, where each node has only three nearest neighbors, leading to strong quantum fluctuations. One of the most exciting discoveries in this realm has been materials that strongly conform with the expectations of the Kitaev model, a model that considers bond-dependent exchange interactions on the honeycomb lattice, as pictured in Fig. \ref{fig:intro_fig}a, which has a quantum spin liquid as its ground state \cite{Kitaev_2006, jakeli} (Fig.~\ref{fig:intro_fig}a). 

In this manuscript, we establish a geometric protocol to identify materials with honeycomb arrangements from large materials databases, particularly the Materials Project \cite{MP1, MP2}. While perfect honeycomb arrangements can be uniquely identified based on space and point group symmetries alone, our filter is more general and can identify materials with distortions from perfect honeycomb arrangements. Moreover, our filter allows us to identify various lower and higher symmetry structural motifs based on hexagonal building blocks, which we term one-dimensional and three-dimensional honeycombs, respectively. In addition to a variety of geometric and chemical descriptors, we also introduce a number of metrics that can be used to narrow down candidate materials for different physics experiments, including a metric that can identify electronic instabilities based on the density-of-states and a cleavability metric based on expected van der Waals bond distances. This approach can be straightforwardly extended to any desired lattice type to expedite the discovery of novel materials.

\section{Geometric Filters}
\label{sec:predictive}

The starting point of this work is large materials databases including those that are purely experimental, such as the Inorganic Crystal Structure Database (ICSD) \cite{ICSD} and the Pearson database \cite{Pearsons} or databases that incorporate computationally predicted materials such as Materials Project \cite{MP1, MP2} or AFLOW \cite{AFLOW}. In this work, we employ Materials Project due to its straightforward APIs enabling rapid access to structures and foundation in \texttt{pymatgen} \cite{pymatgen} syntax, but this approach can be extended to any database.

\begin{figure*}
    \centering
    \includegraphics[width=0.75\linewidth]{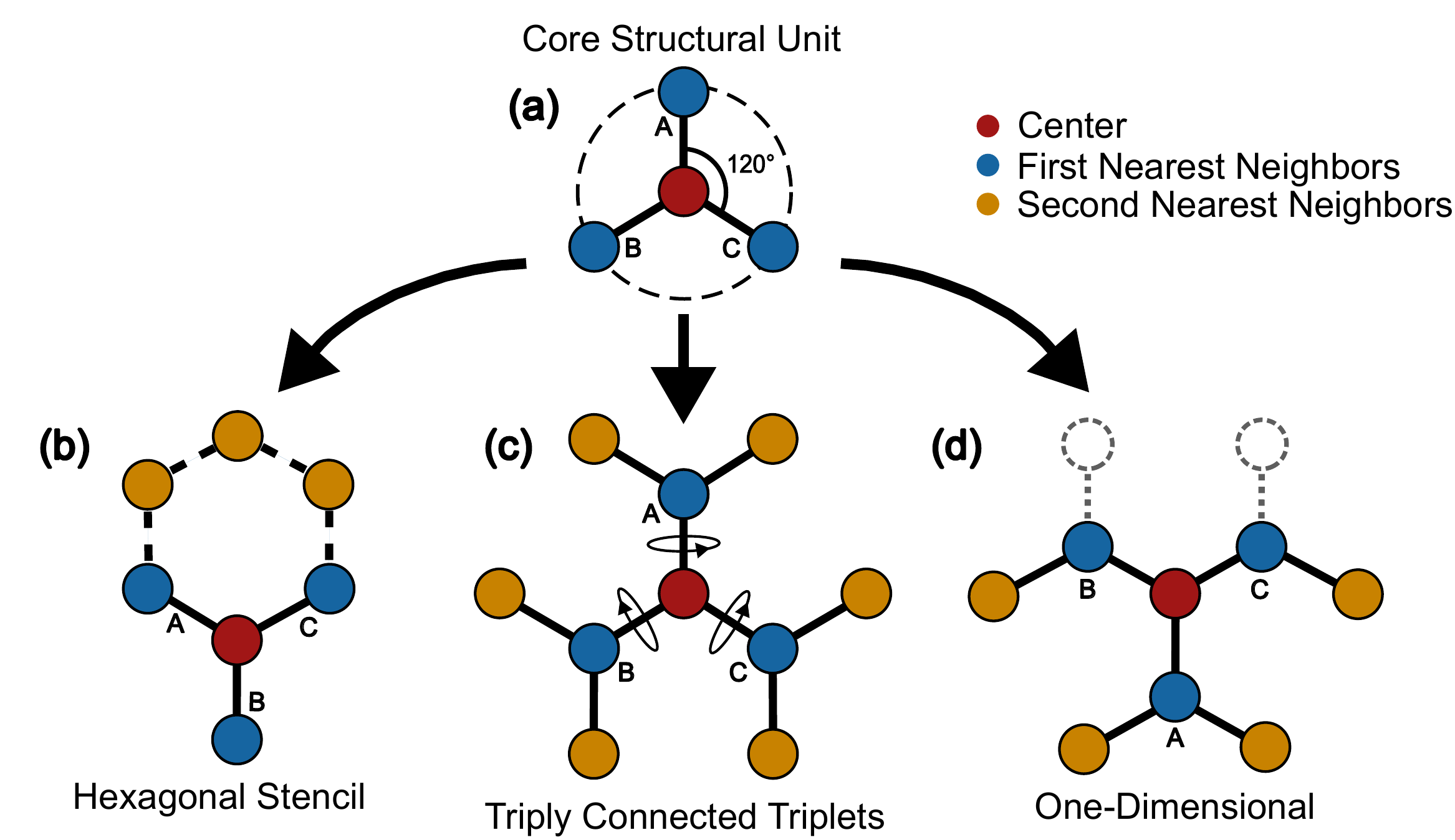}
    \caption{Geometric workflows of each filter. (a) The core of the honeycomb filters is identical in each case: a central site with exactly three nearest neighbors, separated by 120$^\circ$, is found. The three nearest neighbors are labeled A, B, and C. (b) The hexagonal stencil honeycomb filter uses a set of two nearest neighbors and the central site from the core sites to extrapolate the positions of the other three vertices in the hexagon unit. The number of triplets required to have a honeycomb is tunable by the user, but the majority of honeycomb-exhibiting materials are found by choosing one triplet. (c) The triply connected triplets filter treats each of the core nearest neighbors as the center for another core set. These external triplets are not restricted to be coplanar with the original core set, which enables identification of hyperhoneycombs and other higher-dimensional honeycomb structures. (d) The 1D honeycomb filter requires that two of the core nearest neighbors have exactly two nearest neighbors, separated by 120$^\circ$, and one of the core nearest neighbors has exactly three nearest neighbors, also separated by 120$^\circ$. This filter is inclusive of 1D chains, but it is not completely exclusive of certain 2D honeycombs, so operating the TCT filter on its output is required to clean the dataset.}
    \label{fig:filter_logic}
\end{figure*}

The defining structural unit of a honeycomb lattice is a node that has a planar connectivity to three nearest neighbor sites that are spaced at 120 degree intervals, labelled by A, B and C in Fig.~\ref{fig:filter_logic}a. This core structural unit constitutes the basis of most of the filters constructed in this work. 
We initially attempted to identify honeycomb materials by extrapolating hexagons based on the geometry of the defining unit. We iterate through all chemical species within a 3$\times$3 padded supercell and identify structures matching the structural unit. The filter is run on all elements in a structure independent of whether they share a crystallographic site. We use a nearest neighbor distance tolerance of 10\% and an angular tolerance of 10$^\circ$ to ensure that distortions are included in the dataset. The extrapolated sites are compared to the real crystal structure and if the average deviation falls below a specified tolerance, the material is recorded as a honeycomb. 

The hexagonal stencil honeycomb filter has significant drawbacks. It is unable to adequately capture distorted lattices, even with generous tolerances. If a honeycomb is strained away from the core unit (an elongation), the error from the predicted honeycomb will be very large. Similar effects exist for buckling and shear distortions (Sec. \ref{sec:distortions}). For this reason, we do not consider the hexagonal stencil honeycomb filter to be rigorous enough to provide accurate analysis of large numbers of materials.

    We elected instead to identify honeycombs from a geometric stencil, denoted triply connected triplets (TCT) (Fig. \ref{fig:filter_logic}c). Starting from the same core structural unit of a node with three neighbors psaced at 120 degree intervals, we next extrapolate the positions of the second nearest neighbors based on the positions of the first nearest neighbors. For instance, the site A is treated as the center of a new core structural unit and the nearest neighbors are computed. This process is iterated for sites A, B, and C. We elected to have no enforcement of a particular plane that each triplet must reside within, so long as they are locally coplanar. For example, the center site must be coplanar with A, B, and C, but the planes defined by A, B, C and their respective second nearest neighbors need not be parallel to the central lattice plane. Four distinct classes of materials arise from this lack of global coplanarity between the triplets, which we investigate in Section \ref{sec:results}.

One set of honeycomb-like geometry that cannot be identified by the previous filters are one-dimensional lattices: hexagons of sites connected by one edge in one dimension. We use a similar method to the TCT filter to identify these structures. The core structural unit is identified in a lattice and second nearest neighbors are identified identically to the TCT filter. In this case however, sites B and C are only allowed to have two nearest neighbors, with one being the original center site (Fig. \ref{fig:filter_logic}d). This enforces a strict boundary on the honeycomb lattice, enabling us to capture one-dimensional honeycomb motifs.

\section{Data Curation}
\label{sec:limitations}
The effectiveness of any filters we construct are ultimately limited by the quality and quantity of source material. We elected to use the Materials Project for our structure data due to the user-friendly API and out-of-the-box integration with \texttt{pymatgen} \cite{pymatgen} for further analysis. The Materials Project, however, suffers from several limitations in data breadth and quality, which we will discuss here.

We identify inconsistencies in some structures, particularly materials that are both observed experimentally and express very high energy above hull (EAH) values. The highest EAH in the TCT database, considering only observed materials, is Co$_7$W$_6$ (mp-1105377) with an EAH of 2.27 eV/atom. We identify the origin of this unphysically large EAH as a mistabulated c-axis lattice parameter, which is double that of the original report of the material \cite{Co7W6_orig}. We have found dozens of similar examples in the Materials Project dataset when sorting by EAH. More concerning are those materials which have incorrect structures, but are not so unphysical that they are calculated to be unstable by the density functional theory (DFT) used by the Materials Project. These structures may be extremely difficult to screen for.

\begin{figure*}
    \centering
    \includegraphics[width=1\linewidth]{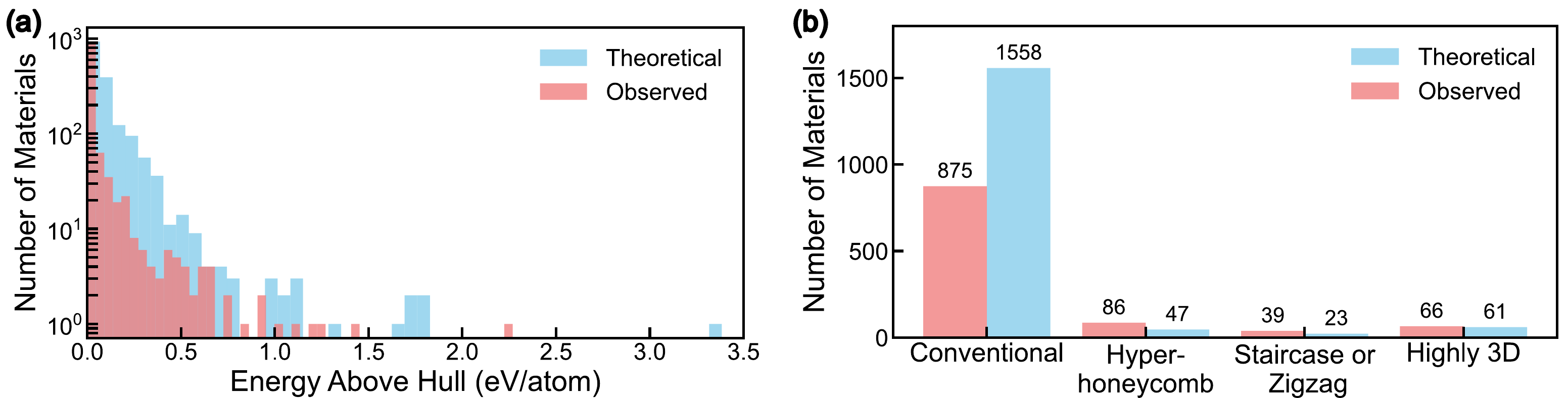}
    \caption{(a) Distribution of the energy above hull values (EAH) for all entries in the TCT database. The experimentally observed materials should have calculated EAH values close to zero. Instead we see many materials with high EAH values, indicating potential issues with either the calculation of the EAH or with the tabulated crystal structure. (b) Distribution of the number of non-coplanar triplets in the TCT database. We denote 0 non-coplanar triplets as conventional honeycombs, one as hyperhoneycombs, two as staircase or zig-zag honeycombs, and three as heavily 3D honeycombs. This classification of the dimensionality of the honeycomb-like lattice is used for all subsequent analysis. }
    \label{fig:narrow_scope}
\end{figure*}

Within the TCT database, the theoretically predicted materials that have not been experimentally observed have systematically larger EAH values than the subset of materials that have been experimentally observed (Fig. \ref{fig:narrow_scope})b. While some of these may be due to issues with stability calculations, the number of observed honeycombs with EAH values above 0.1 eV/atom is more than 10\% of all observed materials in the database. 

The Materials Project also contains a number of duplicate materials, most prominently in the TCT database is the Al(OH)$_3$ series (36 entries). Many of these entries are identical or very similar to one another, with very minor variations in lattice parameters. In order to prevent these duplicates from disproportionately impacting our analysis,
we elect to remove all duplicate entries and retain only the highest symmetry structure. Both of the databases, with duplicates and without, are freely available for subsequent analysis.

\section{Database Composition}
\label{sec:results}
The TCT filter is able to operate directly on .cif files, so an expansive search for honeycombs in larger databases like the Inorganic Crystal Structure Database (ICSD) \cite{ICSD} would be simple in application. We opted for analysis of Materials Project data due to the open-access nature of the database and its large collection of over 150,000 materials. We perform our analysis on the TCT results because the filter can identify and label more exotic quasi-honeycomb systems. The one-dimensional honeycombs are discussed separately.

The TCT filter is capable of identifying conventional coplanar honeycombs, but it is not restricted from finding higher-dimensional pseudo-honeycomb materials. Thus, classifying the dimension of the honeycomb is crucial to classify the materials output by the TCT filter. We use a simple algorithm that identifies the orthogonal vector for each of the external triplets and compares them to the average orthogonal vector of the core structural unit. We use a generous tolerance of 10$^\circ$ to ensure that highly distorted but still two-dimensional honeycombs are classified as such. By identifying how many of the three second-nearest-neighbor triplets are non-coplanar with the core triplets, we assign each material into four categories: zero non-coplanar (conventional 2D honeycombs with potential distortion), one non-coplanar (most commonly hyperhoneycombs, where one triplet is oriented orthogonal to the rest), two non-coplanar (staircase or zigzag honeycombs), three non-coplanar (fully 3D honeycombs, sites making up these lattices are usually far apart and unlikely to contribute to the overall physics of the material in any meaningful way). 

The distribution of non-coplanar triplets, shown in Figure \ref{fig:narrow_scope}a, indicates that conventional honeycombs are by far the most common, although the number of experimentally observed materials is nearly half of the theoretical materials. However, this trend does not exist for the higher-dimensional honeycombs.

\begin{figure*}
    \centering
    \includegraphics[width=1.0\linewidth]{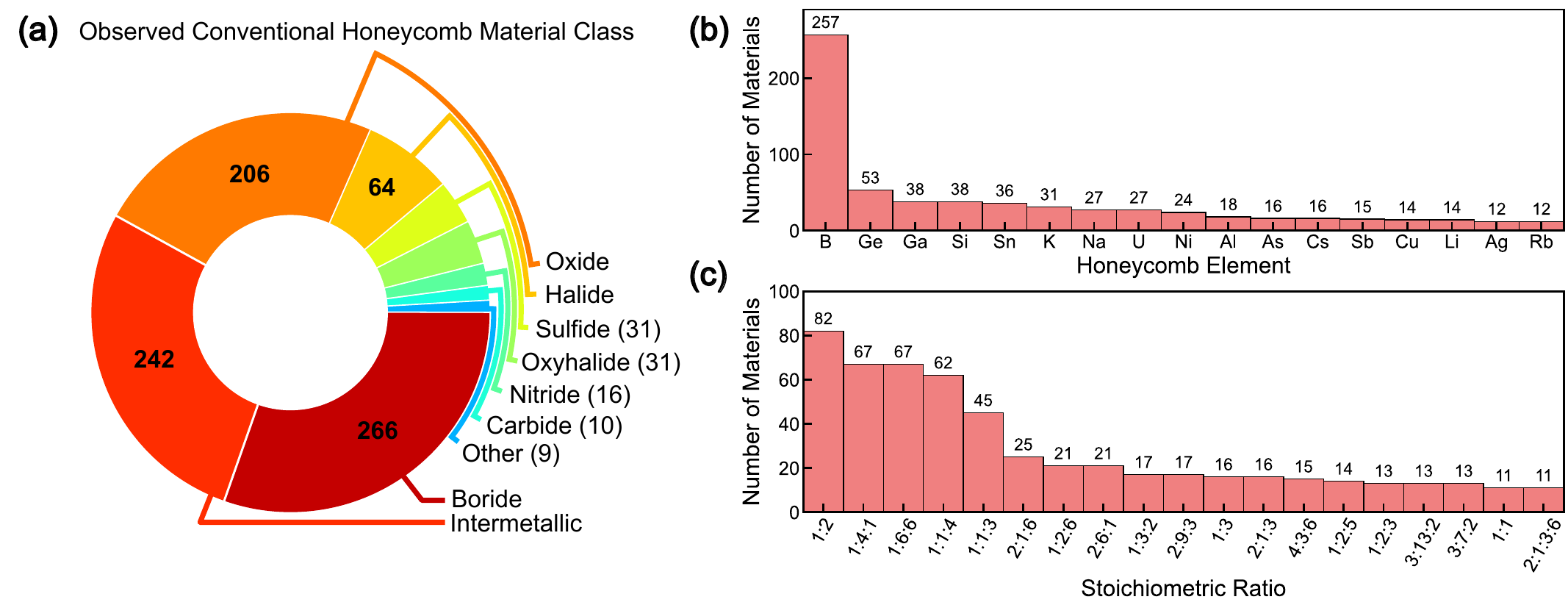}
    \caption{(a) Distribution of material classes within the experimentally observed conventional honeycombs. The classes are defined by a simple sorting process to identify elements present in the chemical formula of each material. For example, materials that contain only intermetallic elements and oxygen are classified as oxides. The number of entries in each category are shown in the label when the pie slice is small. (b) Recorded distribution of honeycomb species across the observed conventional honeycombs. We note that boron is the most common element that makes up honeycombs or quasi-honeycombs in all dimensionality classes. Elements with less than 10 occurrences have been truncated. (c) Stoichiometric ratios defined by the standard chemical formula ordering on Materials Project for the observed conventional honeycombs. Elements are listed from electropositive to electronegative. Because of this ordering, stoichiometric ratios such as 1:4:1 and 1:1:4 are structurally distinct and kept as separate categories. Ratios with less than 10 occurrences have been truncated.}
    \label{fig:conv_honeycombs}
\end{figure*}
\subsection{Conventional Honeycombs}

 The conventional honeycomb class was further categorized according to three attributes: material class, honeycomb element, and stoichiometric ratio (Fig. \ref{fig:conv_honeycombs}). We define material class based on a descending filter of species present in the chemical formula: hydroxides, oxyhalides, halides, oxides, nitrides, sulfides, phosphides, non-metals (graphite), carbides, borides, and intermetallics. The borides are the most common class within the observed conventional honeycombs (Fig. \ref{fig:conv_honeycombs})a, corroborated by the large number of boron honeycombs (Fig. \ref{fig:conv_honeycombs})b. Furthermore, materials with boron honeycombs are most likely to have a stoichiometric ratio of 1:2, 1:4:1, 1:1:4, 2:1:6, 1:2:6, or 2:6:1.

The most commonly observed conventional honeycombs that do not contain boron are the well known 1:6:6 intermetallics \cite{166_structure}, with honeycomb species of Ge and Sn. The remaining intermetallics are distributed through other structure types, with the 1:2 stoichiometric ratio being the second most common occurrence.


\subsection{Distortions}
\label{sec:distortions}
The main benefit of using geometric analysis of structures over symmetry arguments is the tolerance of large distortions. Most work is done on conventional honeycombs due to the variety of theoretical models that exist and the simplicity of the lattice. Many of these theoretical models rely on perfect honeycomb nets, which are not present in all materials with conventional honeycombs.

\begin{figure*}
    \centering
    \includegraphics[width=0.92\linewidth]{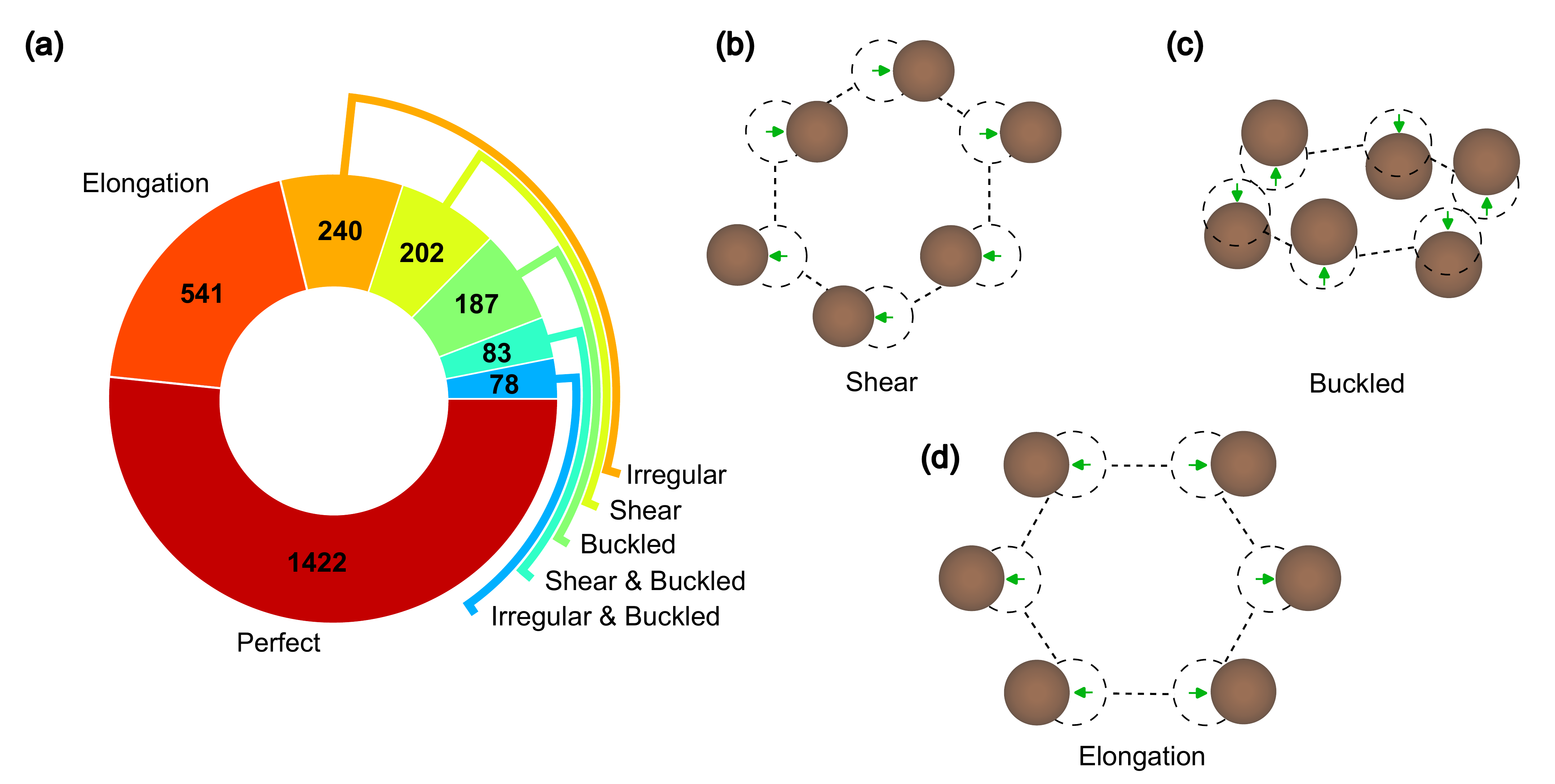}
    \caption{(a) Distribution of distortions in the TCT database within the zero non-coplanar triplets category. Irregular distortions are elongations, contractions, or shear distortions that occur on more than one axis within the honeycomb net. Shears can occur with both elongations or contractions perpendicular to the shear axis, which are compiled into a single category here for ease of viewing. Perfect honeycombs are those in which the planar deviation does not exceed 5\% and the bond length deviations do not exceed 1\%. The number of entries in each category are shown in the label when the pie slice is small. We identify three possible distortion motifs that conventional honeycombs are able to undergo. These are elongations or contractions (d), shear distortions (with elongated or contracted shear axes) (b), or planar buckling (c). On any given material, multiple distortion types may be present.}
    \label{fig:distortion}
\end{figure*}

We identify three general forms of distortions that are possible in conventional honeycombs: elongations (or contractions), shear, and buckling (Fig. \ref{fig:distortion}a-c). Elongations occur when pairs of parallel bonds in a single honeycomb cell are stretched or compressed, while leaving the other bonds the same length. Shear distortions are similar to elongations, but the two halves of the honeycomb cell split by the elongation are distorted in opposite directions along the axis perpendicular to the elongation. Finally, buckling is marked by all sites in the honeycomb net being displaced perpendicular to their lattice plane, most often alternating above and below the plane. Buckling can occur alongside elongations, contractions, or shear distortions.

Distinguishing between a strictly axial length distortion and a shear distortion involves identifying the set of two
bonds that are longer or shorter than the other four (Fig \ref{fig:distortion}a-b). 

Shear distortions can also occur even if there is no variation in bond length. In some cases, distortion along more than one axis is identified, which we define as an ``irregular'' distortion. Buckling (Fig \ref{fig:distortion}c) is a simpler task; if the average displacement from the plane is greater than 5\% of the minimum bond length in the honeycomb cell, we consider the lattice to be buckled. The distribution of all distortions within the TCT database is shown in Figure \ref{fig:distortion}d. As expected, the perfect honeycombs are the most represented, but nearly half of all honeycombs contain major distortions.

The majority of the perfect honeycombs reside in the predictable hexagonal symmetries, which would be collected by a simpler symmetry-oriented filtering process. However, there are some notable exceptions: the second most populous space group after the dominant P6/mmm classification is C2/m, which is a low symmetry monoclinic space group that does not nominally include perfect honeycombs. On more careful examination, we find that these materials indeed include perfect honeycombs, but the symmetry is distorted by other structures, usually bridging ligands.

In contrast, none of the highest represented space groups in the elongated honeycombs are hexagonal. In this set, nearly all of the space groups are very low symmetry, with the highest populations in C2/m, Cmcm, P1, and P$\bar{1}$. This trend is very similar for the remaining distortion categories, with the notable exception of the buckled honeycombs. When buckling is the only distortion, we recover a higher population of hexagonal space groups, with R$\bar{3}$, P$\bar{3}$m1, R$\bar{3}$m, R3m as the most common. 

\subsection{Non-Conventional Honeycombs}

Moving beyond the conventional honeycombs, the TCT filter is able to identify honeycomb-like lattices in arbitrary dimension. Some materials categorized with higher dimension honeycombs, specifically those with three non-coplanar triplets, are unlikely to contribute significantly to the physical properties of the materials because the sites making up the honeycomb are widely separated. Nevertheless, the structural motifs exist within these materials, so we include them for completeness.

\begin{figure*}
    \centering
    \includegraphics[width=1\linewidth]{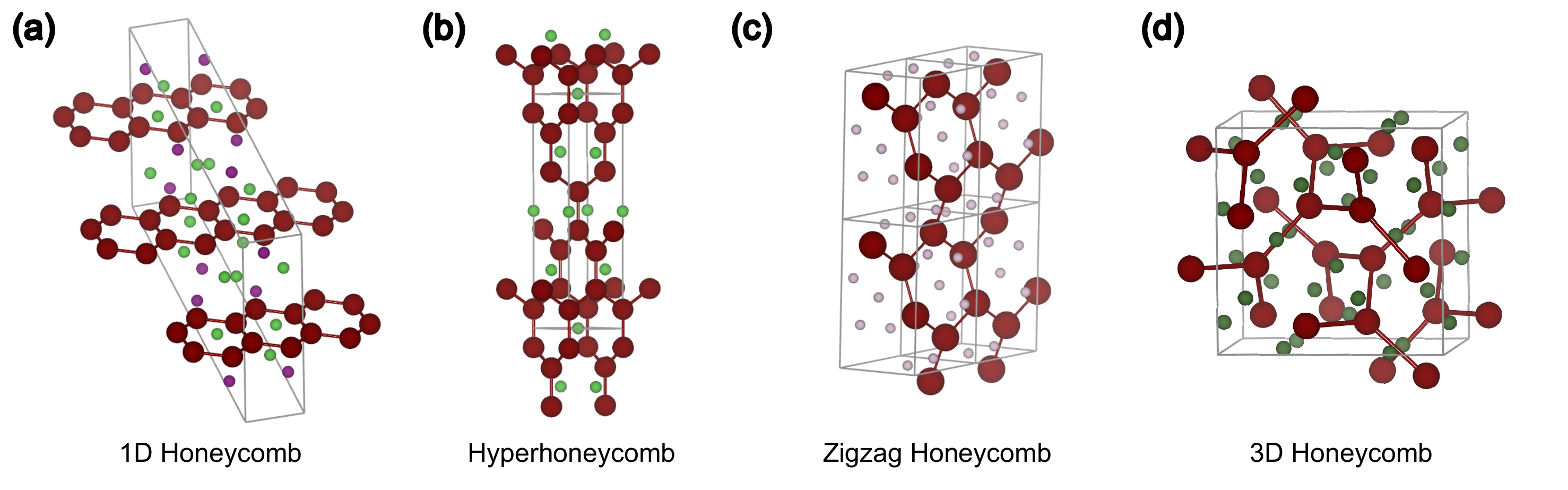}
    \caption{Examples of materials exhibiting different honeycomb-like motifs. (a) La$_3$(AlI)$_2$ (mp-29987) with the Al honeycomb chains highlighted in brown. The chains are separated by La (green) sites. (b) LaGe$_2$ (mp-19761) with the Ge sites highlighted in brown. The Ge sites making up the honeycomb-like structure have two coplanar triplets and one rotated by $90^\circ$. This structure is often referred to as a hyperhoneycomb \cite{hyperhoneycomb}. (c) Re$_2$(HgO$_2$)$_5$ (mp-555759) is an example of materials with extreme buckling in the honeycomb lattice (shown in brown). The oxygen sites were removed for clarity. The distortion in the lattice is so severe that the material was registered with two non-coplanar triplets by the TCT filter. The two non-coplanar triplets category has the largest diversity of honeycomb-like motifs, another being staircase honeycombs. (d) Er$_3$Ru$_2$ (mp-1190026) is an example of a materials with three non-coplanar triplets, identified in brown.}
    \label{fig:num_nonplanar}
\end{figure*}

In addition to the TCT filter, we employ a second filter designed to identify one-dimensional honeycomb systems. The filter is inclusive of one-dimensional honeycombs, but some 2D systems also pass the filter under specific conditions. For this reason, we run the TCT filter on the output of the one-dimensional filter to remove the extraneous materials. The final dataset contains 135 materials, the majority of which contain honeycombs made of boron or silicon. Of the small number of intermetallic one-dimensional honeycombs, we identify two interesting cases: La$_3$(AlI)$_2$ \cite{La3Al2I2_cite} (mp-29987) and Ce$_3$(AlI)$_2$ \cite{Ce3Al2I2_cite} (mp-636773). The La-containing material is shown in Figure \ref{fig:num_nonplanar}a, with the Al honeycomb chains highlighted in brown.

Returning to the results of the TCT filter, we identify three structure types beyond the conventional honeycombs defined by zero non-coplanar triplets. Structures exhibiting one non-coplanar triplet are predominantly hyperhoneycombs, which are a known structural motif and purported to host interesting physics \cite{hyperhoneycomb_magnetism, hyperhoneycomb_transport}, such as a theorized 3D quantum Hall effect \cite{hyperhoneycomb_qhe}. Relative to the conventional honeycombs, the number of one non-coplanar materials in the database is small, with only 86 observed entries, but is the highest number of entries excluding the conventional honeycombs. An example of an non-distorted hyperhoneycomb is LaGe$_2$ (mp-19761) (Fig. \ref{fig:num_nonplanar}b), with the honeycomb made up of Ge sites. Hyperhoneycombs are characterized by the non-coplanar triplet being rotated perpendicular to the others, but the TCT filter will capture the material regardless of this angle. For this reason, the one non-coplanar triplets dataset contains perfect hyperhoneycombs, distorted hyperhoneycombs, and other, less common, types of quasi-honeycomb materials.

The most inconsistent set of materials in the database are the two non-coplanar triplet materials. There are a large variety of structures that adhere to the two non-coplanar triplet condition, such as staircase, zigzag, and larger buckled honeycombs. An example of an extended buckled honeycomb is Re$_2$(HgO$_2$)$_5$ (mp-555759) (Fig. \ref{fig:num_nonplanar}c). In the two non-coplanar triplets case, the honeycomb cells are complete and planar, with no distortions or minor elongations or contractions. The buckling occurs on a much larger length scale than in the conventional honeycombs, which results in different structural motifs.

The set of materials with three non-coplanar triplets are more consistent in structure than the two non-coplanar materials. A representative example of three non-coplanar triplets is Y$_3$BBr$_3$ (mp-1207808) (Fig \ref{fig:num_nonplanar}d), with the honeycomb in B. In materials like Y$_3$BBr$_3$, each triplet is rotated by a consistent angle from the plane defined by the central set of points. While this structure type is interesting given the unusual honeycomb structure, it is unlikely that it will contribute to the overall physical properties of the material due to the extremely long distances between honeycomb sites.
Analysis of these higher-dimensional honeycombs are not as common as the conventional 2D honeycombs, so we hope that these classifications incite more robust investigations.
\begin{figure*}
    \centering
    \includegraphics[width=0.9\linewidth]{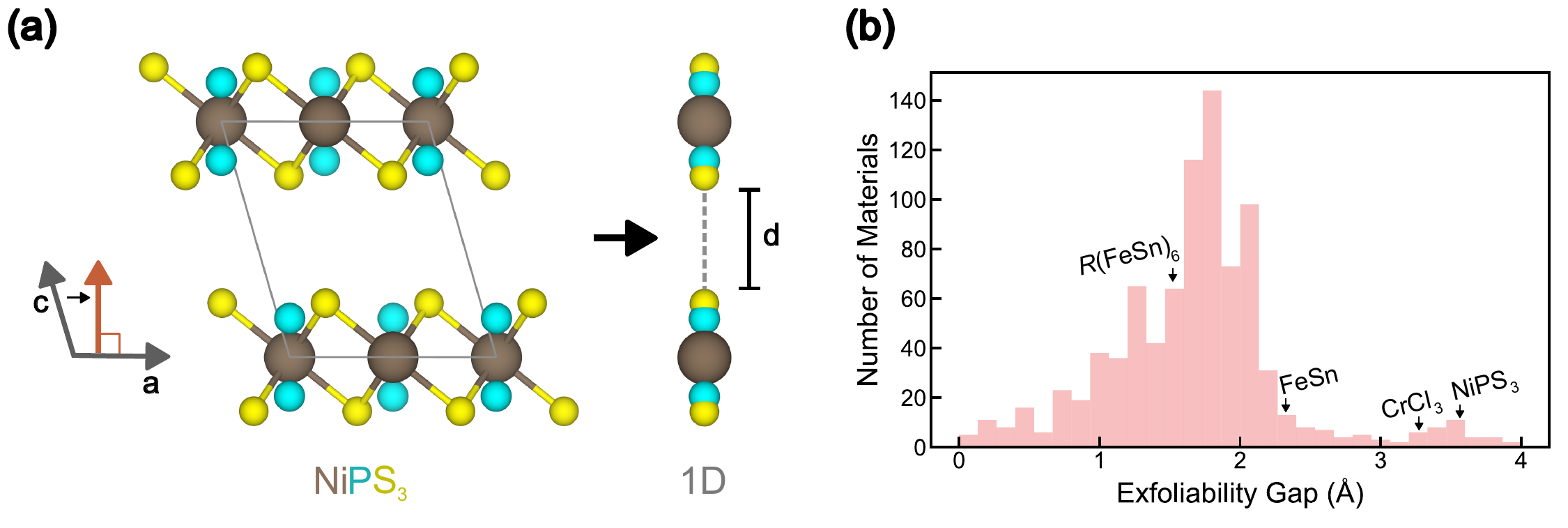}
    \caption{(a) Materials with conventional honeycombs are analyzed by obtaining the orthogonal vector to the honeycomb lattice and projecting all sites in the superstructure onto that one-dimensional axis. The coordinates of the sites are then sorted into ascending order and the largest gap is taken as the exfoliability gap. (b) Distribution of exfoliability gaps in observed conventional honeycomb materials. Two modes are apparent, with the one at >3\space\AA \space roughly representing the exfoliable materials.}
    \label{fig:exfoliability}
\end{figure*}

\section{Physical Property Analysis}

Beyond observing basic structural trends in the identified honeycombs, analysis of the physical properties of the materials, originating from magnetic and electronic properties, are essential for identifying potentially interesting materials. We highlight three such analyses here, exfoliability, density of states analysis, and magnetic analysis. None of these simple models are perfectly deterministic, but we consider them useful for narrowing the scope of the database and creating a tunable interface for identifying materials with targeted properties. 

\subsection{Exfoliability}

In condensed matter physics, cleavability and exfoliability are both highly desirable material attributes. Exfoliability enables materials to be studied in the two-dimensional limit, where electronic properties can shift dramatically from the 3D counterparts. Graphene, for example, is an exfoliated monolayer of graphite. Cleavability on the other hand enables powerful but surface sensitive experimental techniques like angle-resolved photoelectric spectroscopy (ARPES) and scanning-tunneling electron microscopy (STM) to be performed, providing unqiue electronic insights. Here we develop a filter to search for exfoliable materials, for which cleavability can also be expected. The converse, however, is not true. There are many cleavable materials that do not exfoliate to a stable monolayer.
Prediction of cleavable materials requires careful consideration of electronic structure and bonding orbitals around the mirror planes within a material. Conversely, materials that exfoliate into monolayers can be identified by large gaps between 2D structures, called van der Waals gaps (vdW). There are several materials databases that employ computational techniques to more carefully identify exfoliable materials, such as the Materials Cloud Two-Dimensional Structure Database \cite{MC2D_1, MC2D_2} and 2DMatPedia \cite{2DMat}. Here, we employ a simple geometric argument to identify vdW gaps in materials and quantify their sizes. 

We quantify vdW gaps using the orthogonal vector of a 2D honeycomb, which can be returned by any of the filter techniques described previously. A linear transformation is applied to orient the c-axis of the unit cell with the orthogonal vector of the honeycomb lattice. Next, the c-axis cartesian coordinates are taken from each point and sorted in ascending order. The largest gap between subsequent values is the exfoliability gap (Fig \ref{fig:exfoliability}a). Importantly, the range of the sorted list is restricted to the bounds of the original unit cell to prevent miscalculations with flat primitive cells. This functionally expands the edges of the unit cell in-plane without changing the c-axis.

The resulting distribution of exfoliability gaps, specifically for observed materials, is shown in Figure \ref{fig:exfoliability}b. The distribution is bimodal around 1.8\space\AA\space and 3.5\space\AA. The lower mode is characteristic of normal bonding distances between atoms in materials, so the exfoliability gap below 3~\AA\space is not deterministic. Above this limit however, we find that the second mode is comprised of highly 2D layered materials with large gaps. These materials, while not guaranteed to be exfoliable, are significantly more likely to be than those below the 3\space\AA\space threshold. We note that there is a 0.15\space\AA\space separation between the 2.59\space\AA\space exfoliability gap for BaGa$_2$ (mp-1219), which does not have vdW gaps upon inspection, and the 2.74\space\AA\space exfoliability gap for Al(SiO$_3$)$_2$ (mp-732264), which does have an obvious vdW gap. Outside of this separation, the exfoliability gaps vary smoothly, suggesting that $\approx2.7$\space\AA\space may be the exfoliablity threshold.

The exfoliability gap can identify entries in Materials Project data that may contain inconsistencies, much like the EAH analysis discussed in \ref{sec:limitations}. We identify four unreasonably high exfoliability gap materials: Co(RhO$_3$)$_2$ (mp-1226570) with a gap of 8.59\space\AA, Mg$_3$Al(SiO$_4$)$_3$ (mp-1222440) with a gap of 7.86\space\AA, Co$_7$W$_6$ (mp-1105377) with a gap of 6.12\space\AA, and Rb$_2$NbCuS$_2$ (mp-1210898) with a gap of 4.85\space\AA. All of these materials except Co$_7$W$_6$ (which was discussed in detail in Section \ref{sec:limitations}) are theoretical materials with EAH values that are much more reasonable than the extreme values discussed in Section \ref{sec:limitations} (Co(RhO$_3$)$_2$: 0.148 eV/atom, Mg$_3$Al(SiO$_4$)$_3$: 0.168~eV/atom, Rb$_2$NbCuS$_2$: 0.524 eV/atom). This is further demonstration that the EAH analysis is insufficient to identify all unphysical materials in Materials Project, so a multi-pronged approach is necessary.

\subsection{DOS Analysis}
A peak in the density of states (DOS) at the Fermi energy means that many electronic states are available at very low energies, strongly enhancing the material's response to interactions. This can promote instabilities such as itinerant magnetism, superconductivity, or charge-density-wave order, particularly when the peak arises from a van Hove singularity or flat band \cite{flat_bands_fermi}. In order to screen for materials that might exhibit this phenomenology, we develop a filter to identify such peaks in the DOS based on the DFT calculated band structures. Identification of peaks can be done using four descriptive parameters: the height of the peak relative to the background, the quality of the peak (how well the DOS fits a Lorentzian distribution based on relative area), the prominence of the peak (the concentration of electronic states to a particular energy), and how close the center of the peak is to the Fermi energy.
\begin{figure*}
    \centering
    \includegraphics[width=1\linewidth]{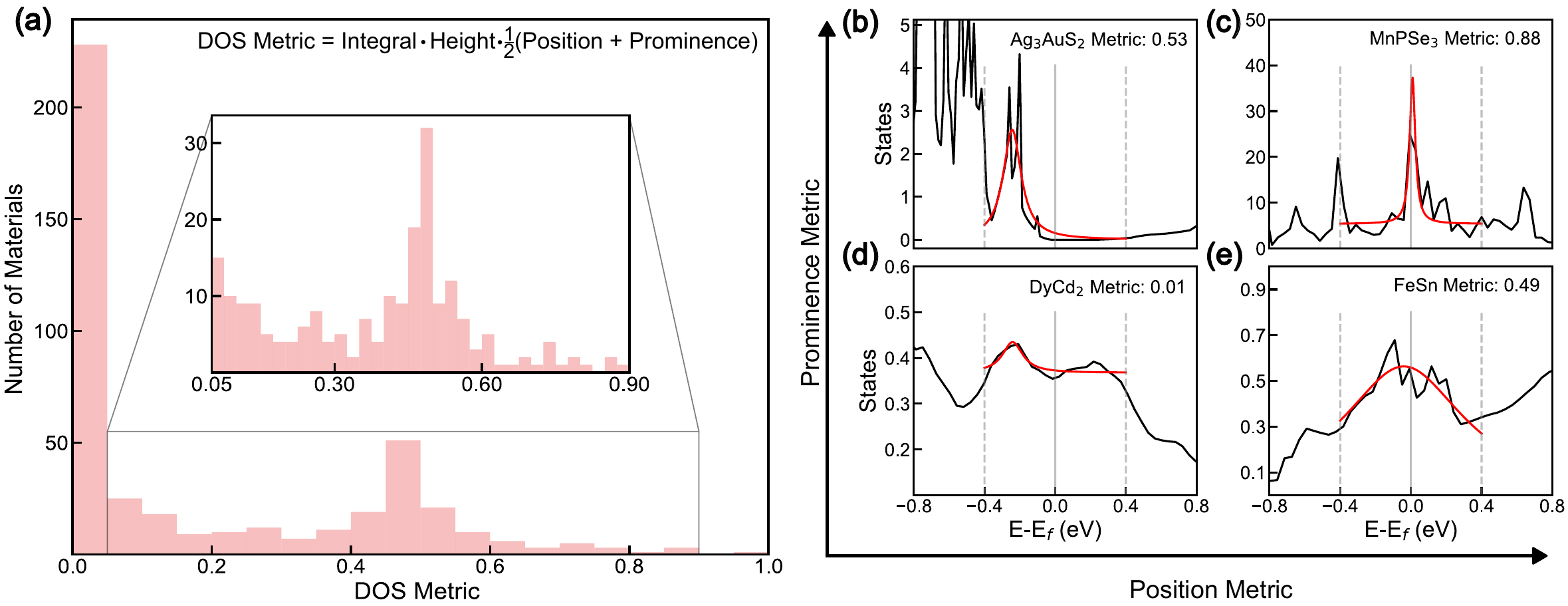}
    \caption{Results of the weighted DOS analysis on experimentally observed honeycomb materials with calculated density of states in the Materials Project. (a) Distribution of the DOS metrics on observed honeycomb materials. The inset is truncated at $M=0.05$ in order to exclude the poorly rated materials represented by the large column around $M=0$ in the histogram. The DOS metrics are normally distributed around $M=0.5$ with a preference for lower DOS metric values. (b-e) Examples of DOS analysis on each ``corner'' of the cross correlation between the DOS sub-metrics. Ag$_3$AuS$_2$ (mp-27554) has high prominence, height, and integral metrics, but a poor position metric. MnPSe$_3$ (mp-8695) has high values in all metrics, while DyCd$_2$ (mp-11294) has low values in all metrics. Finally, FeSn (mp-21260) has the opposite traits to Ag$_3$AuS$_2$: high position metric, but poor prominence.}
    \label{fig:dosanalysis}
\end{figure*}
We first fit the DOS obtained from the Materials Project \cite{MP1, MP2, MP_electstruct} in the vicinity of the Fermi energy (-0.4 eV to +0.4 eV) to a Lorentzian distribution:
\begin{equation}\label{eq:lorentz}
    f(E, a, b, c, d) = \frac{a}{1+(\frac{E-c}{b})^2} + d,
\end{equation}
where $E$ is the energy, $a$ is the height of the distribution, $b$ is the width of the distribution, $c$ is the displacement of the center, and $d$ is the vertical displacement. In order to quantify quality of the DOS peak using these four parameters, we develop the following metrics. The goal is to obtain a single number, from 0 to 1, indicating the quality of the peak in the DOS, where 0 represents no states present in the range, and 1 represents a $\delta$-function at the Fermi energy.

In order to quantify how close the peak is to the Fermi energy, we compute the overlap between the Lorentzian and a copy centered at the Fermi energy. In order to enforce that the resulting position metric is strictly from 0 to 1, both Lorentzians are restricted to $a=1$ and the overlap integral is normalized:
\begin{equation}\label{eq:position}
    p(b,c,d) = \frac{2}{\pi b}\int_{-\infty}^{\infty}f(E,1,b,c,d)\cdot f(E,1,b,0,d)dE
\end{equation}

The quality of the fit, or integral metric, was determined by comparing the area under the curve of the raw DOS and that of the fit Lorentzian. The computed ratio was then subject to Gaussian distribution around the ideal ratio of 1:
\begin{equation}\label{eq:integral}
    g(R,L) = \exp\Big(\frac{-(\frac{R}{L}-1)^2}{2\cdot \sigma_1^2}\Big),
\end{equation}
where $R$ represents the total area under the DOS and $L$ represents the area under the fitted Lorentzian. The width of the Gaussian was selected in order to ensure that peaks within $\sigma_1=0.2$ eV of the Fermi energy are guaranteed to lie within the first standard deviation.

The prominence metric of the Lorentzian fit is quantified by the width of the Lorentzian $b$, which we subject to a Gaussian distribution to recover a standardized metric. 
\begin{equation}\label{eq:thin}
    k(b)=\exp\Big(\frac{-b^2}{2\cdot \sigma_2^2}\Big)
\end{equation}
The width of the Gaussian was again selected to ensure that peaks with a width of less than $\sigma_2=0.15$ eV are weighted within the first standard deviation of the distribution.

The height of the Lorentzian distribution $a$ was compared to the parameter $d$, which indicates the height of the background. The resulting ratio is then subject to a sigmoid function in order to ensure that the resulting height metric lies between 0 and 1. 
\begin{equation}
    h(a,d)=\frac{1}{\exp(-(\frac{a}{d}-x_0)/w)+1}
\end{equation}
The displacement of $x_0=2.6$ was selected to ensure that the height of the DOS peak is significantly taller than the height of the DOS background, but this value can, and should, be modified if a more specific criteria is desired. The width of $w=0.5$ was selected to ensure a smooth transition from the more desirable height ratio to the less desirable ratio. The smoothing should be roughly 1/5 of the displacement value selected to preserve the same trend.

The final DOS peak metric is calculated from an averaging of the prominence and position metrics, which dictate the overall quality of the DOS peak, followed by a weighting by both the integral and height metrics:
\begin{equation}
    M = g\cdot h\cdot\Big(\frac{p + k}{2}\Big)
\end{equation}
Density of states objects were downloaded via the Materials Project API for analysis. The Materials Project provides orbital-projected DOS, so we are able to perform the DOS peak analysis on the species making up the honeycomb lattice.

The resulting DOS metrics from the experimentally observed honeycomb-containing materials are shown in Figure \ref{fig:dosanalysis}a. Apart from the very low rated DOS peaks, which are dominated by flat DOS or gaps around the Fermi energy, the higher DOS metrics are normally distributed around $M=0.5$. 

The height and integral metrics are concentrated near 0 and 1, since they are meant only to weigh out DOS with no significant peaks or poorly fitting Lorentzians. The sensitivity of the height metric and integral metric are what generate the large concentration of materials near $M=0$, since the prominence and position metrics are irrelevant in the case of a poorly fitting Lorentzian or an insignificantly high peak. Thus, the DOS metrics higher than $M\approx0.05$ represent fits that have passed the constraints imposed by the height and integral metrics.

The prominence and position metrics are nearly entirely responsible for the placement of peaks at high DOS metrics. A prominence metric of 0 and a position metric of 1 is functionally identical to the inverse. Since there are two motifs that in the DOS that can result with $M=0.5$, while having a strictly low or high DOS metric requires both of the constraints to be satisfied, it makes sense that there would be a larger concentration of DOS metrics near $M=0.5$, which is observed in Figure \ref{fig:dosanalysis}a.

\subsection{Magnetism}

Honeycomb lattices are famous for hosting interesting magnetic properties due to the frustration that arises from competing exchange interactions. The Kitaev model of magnetism on the honeycomb lattice is one of the only analytically solvable models that predicts quantum spin liquids \cite{Kitaev_2006}, a novel quantum state that is a current subject of intense research. For these reasons, identifying honeycomb materials with magnetic species is useful for determining potential candidates for further research. 

We define two simple identifications of potentially significant magnetic interactions. The first follows from ligand-mediated superexchange interactions described by Goodenough \cite{goodenough1, GOODENOUGH2} and Kanamori \cite{KANAMORI}. Magnetic 3\textit{d} species in the honeycomb net are identified by oxidation states calculated by the \texttt{pymatgen} bond valence analyzer implementation: V$^{3+}$, V$^{4+}$, Cr$^{3+}$, Cr$^{4+}$, Cr$^{2+}$, Mn$^{2+}$, Mn$^{3+}$, Mn$^{4+}$, Mn$^{6+}$, Fe$^{2+}$, Fe$^{3+}$, Co$^{2+}$, Co$^{3+}$, Co$^{4+}$, Ni$^{2+}$, Ni$^{3+}$, and Cu$^{2+}$. If the adjacent honeycomb sites possess bridging ligands, specifically those with one or two donor electrons, including the chalcogens and halogens, we consider the material to be potentially magnetic through the superexchange mechanism. Because of competing energy scales, continuing with 4\textit{d} and 5\textit{d} elements is riskier for false positives than using the simpler 3\textit{d} species.

Lanthanides with unpaired electrons retain their local moments regardless of the chemical environment and therefore are likely candidates for magnetism \cite{lanthanidemoments}, except in rare cases such as the valence instabilities occasionally observed for Ce$^{3+}$Ce$^{4+}$, Eu$^{2+}$/Eu$^{3+}$, and Yb$^{3+}$/Yb$^{2+}$. Following from this, the second method for identifying magnetism first identifies if the honeycomb species is a lanthanide, excluding La and Lu. If the material is both insulating and the rare-earth bond is larger than 5~\AA, the exchange interaction is likely to be so weak that the moments are rendered non-interacting. These simple models are meant to quickly narrow down larger databases of materials to a list of materials to be subject to more rigorous analysis, not to be determinative of magnetic interactions. 

Within the TCT database, we find 497 materials with potential ligand-based superexchange magnetism in the 3\textit{d} honeycomb elements, and 132 interacting 4\textit{f} rare earth honeycombs. 

\section{Conclusion}
In this work, we presented a geometric filtration process for identifying heavily distorted honeycomb lattices in arbitrary crystal structures. The code is functional for application to general CIF files or interfacing with online crystallographic databases, particularly the Materials Project. The TCT filter allowed us to identify several classes of honeycomb-like structures: conventional honeycombs, hyperhoneycombs, staircase or zigzag honeycombs, or heavily 3D honeycombs. The Materials Project database contains mostly conventional honeycombs, as expected, but also contains several hundred higher-dimensional honeycombs. We also discussed a one-dimensional honeycomb filter, which was able to find roughly 100 entries in the Materials Project exhibiting that motif. We demonstrated the analysis of the honeycomb-like materials identified by the TCT filter through several structural, magnetic, and electronic post-processing functions, specifically characterizing the exfoliability, basic magnetic character, and density-of-states peak identification. These filters can enable the accelerated discover of new materials for a wide range of correlated electron problems. These filters can enable the accelerated discovery of new materials for a wide range of correlated electron problems.

\section*{Data availability}
The data that support the findings of this study are available from the corresponding author upon reasonable request.

\section*{Acknowledgments}

This work was supported by the Natural Sciences and Engineering Research Council of Canada (NSERC) and the Canadian Institute for Advanced Research (CIFAR). AMH was supported by the Killam Accelerator Research Fellowship.

\bibliography{refs}

@article{bilayergraphene,
  title = {Unconventional superconductivity in magic-angle graphene superlattices},
  author = {Cao, Y. and Fatemi, V. and Fang, S. and others},
  journal = {Nature},
  volume = {556},
  pages = {43-50},
  numpages = {7},
  year = {2018},
  publisher = {Nature},
  doi = {10.1038/nature26160},
  url = {https://www.nature.com/articles/nature26160#citeas}
}

@article{hales2001honeycomb,
  title={The honeycomb conjecture},
  author={Hales, T.},
  journal={Discrete \& computational geometry},
  volume={25},
  number={1},
  pages={1--22},
  url={https://link.springer.com/article/10.1007/s004540010071},
  year={2001},
  publisher={Springer}
}

@article{ICSD,
author = "Zagorac, D. and M{\"{u}}ller, H. and Ruehl, S. and Zagorac, J. and Rehme, S.",
title = "{Recent developments in the Inorganic Crystal Structure Database: theoretical crystal structure data and related features}",
journal = "Journal of Applied Crystallography",
year = "2019",
volume = "52",
number = "5",
pages = "918--925",
month = "Oct",
doi = {10.1107/S160057671900997X},
url = {https://doi.org/10.1107/S160057671900997X},
}

@article{MP1,
    author = {Horton, M.K. and Huck, P. and Yang, R.X. and others},
    title = {Accelerated data-driven materials science with the Materials Project},
    journal = {Nature Materials},
    year = {2025},
    volume = {24},
    pages = {1522--1532},
    url = {https://doi.org/10.1038/s41563-025-02272-0}
}

@article{MP2,
    author = {Jain, A. and Ong, S. P. and Hautier, G. and Chen, W. and others},
    title = {Commentary: The Materials Project: A materials genome approach to accelerating materials innovation},
    journal = {APL Materials},
    volume = {1},
    number = {1},
    pages = {011002},
    year = {2013},
    month = {07},
    issn = {2166-532X},
    doi = {10.1063/1.4812323},
    url = {https://doi.org/10.1063/1.4812323}
}

@article{MP_electstruct,
    author = {Munro, J. M. and Latimer, K. and Horton, M. K. and others},
    title = {An improved symmetry-based approach to reciprocal space path selection in band structure calculations},
    journal = {npj Computational Materials},
    year = {2020},
    volume = {6},
    number = {112},
    url = {https://doi.org/10.1038/s41524-020-00383-7},
}

@article{Co7W6_orig,
author = {Magneli, A. and Westgren, A.},
title = {Röntgenuntersuchung von Kobalt–Wolframlegierungen},
journal = {Zeitschrift für anorganische und allgemeine Chemie},
volume = {238},
number = {2-3},
pages = {268-272},
url = {https://onlinelibrary.wiley.com/doi/abs/10.1002/zaac.19382380211},
year = {1938}
}

@article{hyperhoneycomb,
author = {Li, M. and Rousochatzakis, I. and Perkins, N.},
year = {2020},
month = {08},
pages = {},
title = {Reentrant incommensurate order and anomalous magnetic torque in the Kitaev magnet $\beta\text{\ensuremath{-}}\mathrm{Li}_{2}\mathrm{Ir}\mathrm{O}_{3}$},
volume = {2},
journal = {Physical Review Research},
url = {https://doi.org/10.1103/PhysRevResearch.2.033328}
}

@article{La3Al2I2_cite,
author = {Mattausch, H. and Oeckler, O. and Zheng, C.},
year = {2001},
month = {07},
pages = {1523-1531},
title = {Kondensierte {Al}$_6$-Ringe in den Subiodiden $\mathrm{La}_{3}\mathrm{Al}_{2}\mathrm{I}_{2}$ und $\mathrm{La}_{2}\mathrm{Al}_{2}\mathrm{I}$},
volume = {627},
journal = {Zeitschrift F\"ur Anorganische Und Allgemeine Chemie},
doi = {10.1002/1521-3749(200107)627:73.3.CO;2-4}
}

@article{Ce3Al2I2_cite,
author = {Mattausch, H.},
year = {2014},
month = {02},
pages = {376-376},
title = {Crystal structure of tricerium dialuminide diiodide, $\mathrm{Ce}_{3}\mathrm{Al}_{2}\mathrm{I}_{2}$},
volume = {218},
journal = {Zeitschrift f\"ur Kristallographie. New crystal structures},
doi = {10.1524/ncrs.2003.218.4.376}
}

@article{KANAMORI,
title = {Superexchange interaction and symmetry properties of electron orbitals},
journal = {Journal of Physics and Chemistry of Solids},
volume = {10},
number = {2},
pages = {87-98},
year = {1959},
issn = {0022-3697},
doi = {https://doi.org/10.1016/0022-3697(59)90061-7},
url = {https://www.sciencedirect.com/science/article/pii/0022369759900617},
author = {Kanamori, J.},
}

@misc{pymatgen,
author = {Ong, S. P. and Richards, W. D. and Jain, A. and Hautier, G. and Kocher, M. and others},
doi = {10.1016/j.commatsci.2012.10.028},
month = {6},
title = {Python Materials Genomics (pymatgen): A robust, open-source python library for materials analysis},
url = {https://github.com/materialsproject/pymatgen},
year = {2013}
}

@article{
graphene_original,
author = {Novoselov, K. S. and Geim,  A. K. and Morozov, S. V.  and others},
title = {Electric Field Effect in Atomically Thin Carbon Films},
journal = {Science},
volume = {306},
number = {5696},
pages = {666-669},
year = {2004},
doi = {10.1126/science.1102896},
URL = {https://www.science.org/doi/abs/10.1126/science.1102896},
}

@article{dirac_fermions,
author={Novoselov, K. S.
and Geim, A. K.
and Morozov, S. V.
and Jiang, D.
and others},
title={Two-dimensional gas of massless Dirac fermions in graphene},
journal={Nature},
year={2005},
month={Nov},
day={01},
volume={438},
number={7065},
pages={197-200},
issn={1476-4687},
doi={10.1038/nature04233},
url={https://doi.org/10.1038/nature04233}
}

@article{quantum_hall_effect,
  title = {Unconventional Integer Quantum Hall Effect in Graphene},
  author = {Gusynin, V. P. and Sharapov, S. G.},
  journal = {Physical Review Letters},
  volume = {95},
  issue = {14},
  pages = {146801},
  numpages = {4},
  year = {2005},
  month = {Sep},
  publisher = {American Physical Society},
  doi = {10.1103/PhysRevLett.95.146801},
  url = {https://link.aps.org/doi/10.1103/PhysRevLett.95.146801}
}

@article{graphene_mech,
    author = {Sun, Y. W. and Papageorgiou, D. G. and Humphreys, C. J. and Dunstan, D. J. and others},
    title = {Mechanical properties of graphene},
    journal = {Applied Physics Reviews},
    volume = {8},
    number = {2},
    pages = {021310},
    year = {2021},
    month = {04},
    issn = {1931-9401},
    doi = {10.1063/5.0040578},
    url = {https://doi.org/10.1063/5.0040578},
}

@Article{graphene_flat,
author={Lisi, S.
and Lu, X.
and Benschop, T.
and de Jong, T. A.
and others},
title={Observation of flat bands in twisted bilayer graphene},
journal={Nature Physics},
year={2021},
month={Feb},
day={01},
volume={17},
number={2},
pages={189-193},
issn={1745-2481},
doi={10.1038/s41567-020-01041-x},
url={https://doi.org/10.1038/s41567-020-01041-x}
}

@article{graphene_correlated,
  title = {Correlated Insulating States in Twisted Double Bilayer Graphene},
  author = {Burg, G. W. and Zhu, J. and Taniguchi, T. and Watanabe, K. and  others},
  journal = {Physical Review Letters},
  volume = {123},
  issue = {19},
  pages = {197702},
  numpages = {5},
  year = {2019},
  month = {Nov},
  publisher = {American Physical Society},
  doi = {10.1103/PhysRevLett.123.197702},
  url = {https://link.aps.org/doi/10.1103/PhysRevLett.123.197702}
}

@Article{graphene_topology,
author={Choi, Y.
and Kim, H.
and Peng, Y.
and Thomson, A.
and Lewandowski, C.
and others},
title={Correlation-driven topological phases in magic-angle twisted bilayer graphene},
journal={Nature},
year={2021},
month={Jan},
day={01},
volume={589},
number={7843},
pages={536-541},
issn={1476-4687},
doi={10.1038/s41586-020-03159-7},
url={https://doi.org/10.1038/s41586-020-03159-7}
}

@article{honeycomb_stronginteraction,
title = {Competing ferromagnetic and antiferromagnetic phases on the frustrated Ising honeycomb lattice},
journal = {Physica A: Statistical Mechanics and its Applications},
volume = {686},
pages = {131321},
year = {2026},
issn = {0378-4371},
doi = {https://doi.org/10.1016/j.physa.2026.131321},
url = {https://www.sciencedirect.com/science/article/pii/S0378437126000579},
author = {Dias, P. F. and Zimmer, F.M. and Fytas, N.G.  and Schmidt, M.},
}

@article{Kitaev_2006,
   title={Anyons in an exactly solved model and beyond},
   volume={321},
   ISSN={0003-4916},
   url={http://dx.doi.org/10.1016/j.aop.2005.10.005},
   DOI={10.1016/j.aop.2005.10.005},
   number={1},
   journal={Annals of Physics},
   publisher={Elsevier BV},
   author={Kitaev, A.},
   year={2006},
   month=Jan, pages={2–111} }

@article{jakeli,
  title = {Kitaev-Heisenberg Model on a Honeycomb Lattice: Possible Exotic Phases in Iridium Oxides $A_{2}\mathrm{Ir}_{3}$},
  author = {Chaloupka, J. and Jackeli, G. and Khaliullin, G.},
  journal = {Physical Review Letters},
  volume = {105},
  issue = {2},
  pages = {027204},
  numpages = {4},
  year = {2010},
  month = {Jul},
  publisher = {American Physical Society},
  doi = {10.1103/PhysRevLett.105.027204},
  url = {https://link.aps.org/doi/10.1103/PhysRevLett.105.027204}
}

@article{AFLOW,
title = {AFLOWLIB.ORG: A distributed materials properties repository from high-throughput ab initio calculations},
journal = {Computational Materials Science},
volume = {58},
pages = {227-235},
year = {2012},
issn = {0927-0256},
doi = {https://doi.org/10.1016/j.commatsci.2012.02.002},
url = {https://www.sciencedirect.com/science/article/pii/S0927025612000687},
author = {Curtarolo, S. and Setyawan, W. and Wang, S. and Xue, J. and Yang, K. and others},
}

@misc{Pearsons,
  title = {Pearson's Crystal Data: Crystal Structure Database for Inorganic Compounds},
  author = {Villars, P. and Cenzual, K.},
  publisher = {ASM International},
  address = {Materials Park, Ohio, USA},
  note = {Release 2024/25},
  url={https://www.crystalimpact.com/Default.htm}
}

@article{flat_bands_fermi,
  author = {Geng, S. and Wang, X. and Guo, R. and Qiu, C. and Chen, F. and others},
  title = {Experimental realization of dice-lattice flat band at the Fermi level in layered electride YCl},
  journal = {Nature Communications},
  year = {2026},
  volume = {17},
  number = {1},
  url = {https://doi.org/10.1038/s41467-026-69049-0},
  issn = {2041-1723},
}

@article{166_structure,
title = {Crystal structures of ternary rare earth-3d transition metal compounds of the {$RT$$_6$Al$_6$} type},
journal = {Journal of the Less Common Metals},
volume = {72},
number = {2},
pages = {241-249},
year = {1980},
issn = {0022-5088},
doi = {https://doi.org/10.1016/0022-5088(80)90143-5},
url = {https://www.sciencedirect.com/science/article/pii/0022508880901435},
author = {Felner, I. },
}

@article{hyperhoneycomb_magnetism,
  title = {Hyperhoneycomb Iridate $\ensuremath{\beta}\text{\ensuremath{-}}\mathrm{Li}_{2}\mathrm{IrO}_{3}$ as a Platform for {Kitaev} Magnetism},
  author = {Takayama, T. and Kato, A. and Dinnebier, R. and Nuss, J. and others},
  journal = {Physical Review Letters},
  volume = {114},
  issue = {7},
  pages = {077202},
  numpages = {5},
  year = {2015},
  month = {Feb},
  publisher = {American Physical Society},
  doi = {10.1103/PhysRevLett.114.077202},
  url = {https://link.aps.org/doi/10.1103/PhysRevLett.114.077202}
}

@article{hyperhoneycomb_transport,
    author = {Verissimo-Alves, M. and Amorim, R. G. and Martins, A. S.},
    title = {Anisotropic Electronic Structure and Transport Properties
of the H‑0 Hyperhoneycomb
Lattice},
    journal = {The Journal of Physical Chemistry C},
    volume = {121},
    number = {3},
    pages = {1928-1933},
    year = {2016},
    month = {12},
    issn = {1932-7447},
    doi = {10.1021/acs.jpcc.6b10336},
    url = {https://doi.org/10.1021/acs.jpcc.6b10336},
}

@article{hyperhoneycomb_qhe,
  title = {Three-dimensional quantum anomalous Hall effect in hyperhoneycomb lattices},
  author = {Kim, S. W. and Seo, K. and Uchoa, B.},
  journal = {Physical Review B},
  volume = {97},
  issue = {20},
  pages = {201101(R)},
  numpages = {5},
  year = {2018},
  month = {May},
  publisher = {American Physical Society},
  doi = {10.1103/PhysRevB.97.201101},
  url = {https://link.aps.org/doi/10.1103/PhysRevB.97.201101}
}

@article{carrier_mobility,
  title = {Approaching the Dirac Point in High-Mobility Multilayer Epitaxial Graphene},
  author = {Orlita, M. and Faugeras, C. and Plochocka, P. and Neugebauer, P. and Martinez, G. and others},
  journal = {Physical Review Letters},
  volume = {101},
  issue = {26},
  pages = {267601},
  numpages = {4},
  year = {2008},
  month = {Dec},
  publisher = {American Physical Society},
  doi = {10.1103/PhysRevLett.101.267601},
  url = {https://link.aps.org/doi/10.1103/PhysRevLett.101.267601}
}

@article{goodenough1,
    author = {Goodenough, J. B.},
    title = {Theory of the Role of Covalence in the Perovskite-Type Manganites},
    journal = {Physical Review},
    year = {1955},
    volume = {100},
    issue = {564},
    url = {https://doi.org/10.1103/PhysRev.100.564}
}

@article{GOODENOUGH2,
title = {An interpretation of the magnetic properties of the perovskite-type mixed crystals $\mathrm{La}_{1\text{\ensuremath{-}}x}\mathrm{Sr}_{x}\mathrm{Co}\mathrm{O}_{3\text{\ensuremath{-}}\lambda}$},
journal = {Journal of Physics and Chemistry of Solids},
volume = {6},
number = {2},
pages = {287-297},
year = {1958},
issn = {0022-3697},
doi = {https://doi.org/10.1016/0022-3697(58)90107-0},
url = {https://www.sciencedirect.com/science/article/pii/0022369758901070},
author = {J. B. Goodenough},

}

@article{MC2D_1,
author={Mounet, N.
and Gibertini, M.
and Schwaller, P.
and Campi, D.
and Merkys, A.
and others},
title={Two-dimensional materials from high-throughput computational exfoliation of experimentally known compounds},
journal={Nature Nanotechnology},
year={2018},
month={Mar},
day={01},
volume={13},
number={3},
pages={246-252},
issn={1748-3395},
doi={10.1038/s41565-017-0035-5},
url={https://doi.org/10.1038/s41565-017-0035-5}
}

@article{MC2D_2,
    author = {Campi, D. and Mounet, N. and Gibertini, M. and Pizzi, G. and Marzari, N.},
    title = {Expansion
of the Materials Cloud 2D Database},
    journal = {ACS Nano},
    volume = {17},
    number = {12},
    pages = {11268-11278},
    year = {2023},
    month = {06},
    issn = {1936-0851},
    doi = {10.1021/acsnano.2c11510},
    url = {https://doi.org/10.1021/acsnano.2c11510},
}

@Article{2DMat,
author={Zhou, J.
and Shen, L.
and Costa, M. D.
and Persson, K. A.
and Ong, S. P.
and others},
title={{2DMatPedia}, an open computational database of two-dimensional materials from top-down and bottom-up approaches},
journal={Scientific Data},
year={2019},
month={Jun},
day={12},
volume={6},
number={1},
pages={86},
issn={2052-4463},
doi={10.1038/s41597-019-0097-3},
url={https://doi.org/10.1038/s41597-019-0097-3}
}

@Article{lanthanidemoments,
author={Mugiraneza, S.
and Hallas, A. M.},
title={Tutorial: a beginner's guide to interpreting magnetic susceptibility data with the Curie-Weiss law},
journal={Communications Physics},
year={2022},
month={Apr},
day={19},
volume={5},
number={1},
pages={95},
issn={2399-3650},
doi={10.1038/s42005-022-00853-y},
url={https://doi.org/10.1038/s42005-022-00853-y}
}

\end{document}